\documentclass{article}

\title{Conjugate Spinor Solution of the Dirac Equation for the Hydrogen Atom}
\author{Robert Ducharme}

\begin{document}
\maketitle

\centerline{151 Fairhills Dr., Ypsilanti, MI 48197}
\centerline{E-mail: ducharme01@comcast.net}

\begin{abstract}
It is shown the central field Dirac equation can be simplified through the use of real conjugate spinors to substitute for the upper and lower components of the bi-spinor eigensolutions. This substitution reduces the Dirac equation for the hydrogen atom to the problem of solving a single second order differential equation similar to a Klein-Gordon equation but containing additional terms to take account of the spin on the electron. The bi-spinor wave functions are readily constructed once the solution is known in terms of the conjugate spinors.  
\end{abstract}

\section{Introduction}
The central field Dirac equation is usually presented as a pair of coupled first order differential equations for the top $\Psi_a$ and bottom $\Psi_b$ two-component spinor parts of the bi-spinor eigenfunctions \cite{DFL, PS}. The purpose of this paper is to obtain an equivalent formulation of the problem based on a decomposition of $\Psi_a$ and $\Psi_b$ into real conjugate spinors $\Phi$ and $\tilde{\Phi}$. For the case of the hydrogen atom this results in a single second order differential equation for $\Phi$ that gives the same energy spectrum as the first order form of the Dirac equation. 

The first order central field Dirac equation is presented in section 2. The two-component spinors $\Phi$ and $\tilde{\Phi}$ are then defined as conjugate linear combinations of the bi-spinor components $\Psi_a$ and $\Psi_b$. It is shown this transformation has a simple inverse so that bi-spinor wave functions are easy to construct once $\Phi$ and $\tilde{\Phi}$ have been determined. It is significant that $\Phi$ and $\tilde{\Phi}$ are pure real functions. Conjugate spinors have definitions similar to complex functions in terms of $\Psi_a$ and $\Psi_b$ but turn out to be real owing to the fact that $\Psi_a$ and $\Psi_b$ are not both pure real.

The second order form of the Dirac equation for the hydrogen atom is derived in section 3. It is interesting that this equation takes the form of a Klein-Gordon equation but with additional terms to account for intrinsic spin in the Hamiltonian for the electron. An expression for $\tilde{\Phi}$ in terms of $\Phi$ is also derived.

In section 4, the exact solution of the second order Dirac equation for the hydrogen atom is obtained using analytical methods. The energy spectrum for hydrogen is calculated as well as both the conjugate spinor wave functions. The Dirac bi-spinor wave functions are determined as transformations of the conjugate spinors wave functions.

\section{Conjugate Spinors}
Consider a single spinning particle of mass $m_0$ at a radial distance $r$ from the source of a central potential field $V(r)$. The Dirac equation determining the bi-spinor wave function $\Psi$ for this system is
\begin{equation} 
\hbar c\gamma_4\gamma_i \frac{\partial \Psi}{\partial x_i} - \imath \hbar \frac{\partial \Psi}{\partial t} + V(r)\Psi + \gamma_4 m_0c^2\Psi = 0
\end{equation}
where $\gamma_i (i=1,2,3)$ and $\gamma_4$ are $4 \times 4$ Dirac matrices, $\hbar$ is the Planck constant divided by $2\pi$ and $c$ is the velocity of light. This can also be written in the form
\begin{eqnarray} \label{eq: central_dirac}
c\vec{\sigma}  \cdot \vec{p} \left( \begin{array}{c}
\Psi_b \\
\Psi_a 
\end{array} \right) =
\left( \begin{array}{cc}
E-V-m_0c^2 & 0 \\
0 & E-V+m_0c^2 
\end{array} \right)  
\left( \begin{array}{c}
\Psi_a \\
\Psi_b 
\end{array} \right)
\end{eqnarray}
where $\vec{\sigma}$ are the Pauli spin matrices, $\vec{p}$ is the 3-momentum of the particle and $E$ is the total energy. It is also helpful to have the relationship
\begin{equation} \label{eq: sigma_p_std}
\vec{\sigma}  \cdot \vec{p} = \frac{\imath}{r}\frac{\vec{\sigma} \cdot \vec{r}}{r} \left( - r \hbar \frac{\partial}{\partial r} + \vec{\sigma} \cdot \vec{L} \right)
\end{equation}
where $\vec{L}$ is the orbital angular momentum. 

The task ahead is to substitute for the upper $\Psi_a$ and lower $\Psi_b$ spinors in eq. (\ref{eq: central_dirac}) using the following conjugate spinors:
\begin{equation} \label{eq: Phi_1}
\Phi = \frac{1}{2\sqrt{m_0c^2+E}}\left(\Psi_a -  \imath \sqrt{\frac{m_0c^2+E}{m_0c^2-E}} \frac{\vec{\sigma} \cdot \vec{r}}{r} \Psi_b \right)
\end{equation}
\begin{equation} \label{eq: Phi_2}
\tilde{\Phi} = \frac{1}{2\sqrt{m_0c^2+E}}\left( \Psi_a + \imath \sqrt{\frac{m_0c^2+E}{m_0c^2-E}} \frac{\vec{\sigma} \cdot \vec{r}}{r} \Psi_b \right)
\end{equation}
These expressions are readily inverted to give
\begin{equation} \label{eq: Psi_a}
\Psi_a =  \sqrt{m_0c^2+E}(\Phi+\tilde{\Phi})
\end{equation}
\begin{equation} \label{eq: Psi_b}
\Psi_b = \imath \sqrt{m_0c^2-E}\frac{\vec{\sigma} \cdot \vec{r}}{r}(\Phi-\tilde{\Phi}) 
\end{equation}
The goal is to obtain differential equations for $\Phi$ and $\tilde{\Phi}$. Clearly, once $\Phi$ and $\tilde{\Phi}$ have been determined the bi-spinor components $\Psi_a$ and $\Psi_b$ follow from eqs. (\ref{eq: Psi_a}) and (\ref{eq: Psi_b}). One note of caution here is that $\Psi_a$ and $\Psi_b$ cannot both be assumed to be pure real functions.

A spin-1/2 particle in a spherically symmetric potential $V(r)$, has spherical harmonic eigenfunctions $Y_{lm}(\theta,\phi)$ to represent orbital angular momentum and the two-component spinors $\chi_{1/2} = (1,0)^t$ and $\chi_{-1/2} = (0,1)^t$ for the spin states. Here, the $Y_{lm}$ eigenfunctions and the $\chi_{\pm 1/2}$ spinors can be combined to form the spherical spinors ${\cal Y}_{\kappa m}$ using Clebsh-Gordon coefficients. Spherical spinors are simultaneously eigenfunctions of the following set of operators:
\begin{equation}
\hat{L}^2{\cal Y}_{\kappa m} = l(l+1)\hbar^2{\cal Y}_{\kappa m}, \quad \hat{L}_z{\cal Y}_{\kappa m} = m\hbar{\cal Y}_{\kappa m}, \quad \hat{J}^2{\cal Y}_{\kappa m} = j(j+1)\hbar^2{\cal Y}_{\kappa m}
\end{equation}
for orbital angular momentum, z-component of angular momentum and total angular momentum respectively. It is further useful to define the operator
\begin{equation} \label{eq: K}
\hat{K} = -\hbar - \vec{\sigma}  \cdot \vec{L}
\end{equation} 
such that $\hat{K}{\cal Y}_{\kappa m}=\kappa\hbar{\cal Y}_{\kappa m}$ and 
\begin{equation} \label{eq: K_sqr}
\hat{K}^2 = \hat{L}^2 + \hbar \vec{\sigma}  \cdot \vec{L} + \hbar^2
\end{equation}
One significant advantage being that the integer $\kappa$ can be used in place of both the $l$ and $j$ quantum numbers.

The upper and lower components in Dirac bi-spinors can be expressed in the product form
\begin{equation}
\Psi_a = f_a(r){\cal Y}_{\kappa m}, \quad \Psi_b = f_b(r){\cal Y}_{-\kappa m}
\end{equation}
where $f_a$ and $f_b$ are both functions of r. Putting these results into eqs. (\ref{eq: Phi_1}) and (\ref{eq: Phi_2}) and making use of the parity operator equation 
\begin{equation}
\frac{\vec{\sigma} \cdot \vec{r}}{r}{\cal Y}_{\kappa m} = -{\cal Y}_{-\kappa m}
\end{equation}
it can be seen that $\Phi$ and $\tilde{\Phi}$ must have the product forms
\begin{equation}
\Phi = g(r){\cal Y}_{\kappa m}, \quad \tilde{\Phi} = \tilde{g}(r){\cal Y}_{\kappa m}
\end{equation}
where $g$ and $\tilde{g}$ are conjugate functions of r. Clearly, the conjugate spinors share a common spherical spinor whereas the upper and lower components of a Dirac bi-spinor do not.

Inserting eqs. (\ref{eq: Psi_a}) and (\ref{eq: Psi_b}) in to eq. (\ref{eq: central_dirac}) gives 
\begin{equation} \label{eq: conj_dirac_fo_1a}
\left( \frac{\partial}{\partial r} + \frac{\hbar - \hat{K}}{\hbar r} \right)(\Phi-\tilde{\Phi}) = \frac{-1}{\hbar c} \left( \lambda + \frac{m_0c^2V}{\lambda}+\frac{EV}{\lambda}\right)(\Phi+\tilde{\Phi})
\end{equation}
\begin{equation} \label{eq: conj_dirac_fo_2a}
\left( \frac{\partial}{\partial r} + \frac{\hbar + \hat{K}}{\hbar r} \right)(\Phi+\tilde{\Phi}) = \frac{-1}{\hbar c} \left( \lambda - \frac{m_0c^2V}{\lambda}+\frac{EV}{\lambda}\right)(\Phi-\tilde{\Phi})
\end{equation}
having made use of eq. (\ref{eq: K}) and set $\lambda=\sqrt{m_0^2c^4-E^2}$. Further manipulation of eqs. (\ref{eq: conj_dirac_fo_1a}) and (\ref{eq: conj_dirac_fo_2a}) leads to
\begin{equation} \label{eq: conj_dirac_fo_1}
\left(\hbar\frac{\partial}{\partial r} + \frac{\lambda}{c} +
\frac{EV}{c \lambda} +\frac{\hbar}{r}\right) \Phi = -\left(\frac{\hat{K}}{r} +
\frac{m_0cV}{\lambda} \right) \tilde{\Phi}
\end{equation}
\begin{equation} \label{eq: conj_dirac_fo_2}
\left(\hbar\frac{\partial}{\partial r} - \frac{\lambda}{c} -
\frac{EV}{c \lambda} + \frac{\hbar}{r}\right) \tilde{\Phi} = -\left(\frac{\hat{K}}{r} -
\frac{m_0cV}{\lambda} \right) \Phi
\end{equation}
Eqs. (\ref{eq: conj_dirac_fo_1}) and (\ref{eq: conj_dirac_fo_2}) constitute the first order central field Dirac equation in terms of complex conjugate spinors. These results will be applied to the hydrogen atom in the next section to derive a second order differential equation for $\Phi$, similar to a Klein-Gordon equation, alongside an auxiliary relationship giving $\tilde{\Phi}$ in terms of $\Phi$.

\section{The Second Order Dirac Equation}
The electrostatic potential for the hydrogen atom is
\begin{equation}
V(r) = -c\hbar\frac{\alpha}{r} 
\end{equation}
where
\begin{equation}
\alpha = \frac{e^2}{4 \pi \epsilon_0 c \hbar} 
\end{equation}
is the fine structure constant, -e is the charge on the electron and $\epsilon_0$ is the permittivity of free space.  Combining eqs. (\ref{eq: conj_dirac_fo_1}) and (\ref{eq: conj_dirac_fo_2}) for this potential leads to the second order differential equation
\begin{equation}
\left(\frac{\hat{K^2}}{r} - \frac{m_0^2c^4\hbar^2\alpha^2}{\lambda^2 r} \right) \Phi = \left( \hbar\frac{\partial}{\partial r} -\frac{\lambda}{c} + \frac{\hbar E \alpha}{\lambda r} + \frac{\hbar}{r} \right)\left( \hbar r\frac{\partial}{\partial r} +\frac{\lambda r}{c} - \frac{\hbar E \alpha}{\lambda } + \hbar \right) \Phi
\end{equation}
Expanding the bracket and simplifying this equation with the help of eqs. (\ref{eq: sigma_p_std}) and (\ref{eq: K_sqr}) gives
\begin{equation} \label{eq: dirac_so}
-c^2\hbar^2 \nabla^2 \Phi + m_0^2c^4\Phi = \left( E + c\hbar \frac{\alpha}{r} \right)^2 \Phi + \Gamma \Phi 
\end{equation}
where
\begin{equation}
\nabla^2 = \frac{1}{r^2} \frac{\partial}{\partial r} \left(r^2\frac{\partial}{\partial r} \right) - \frac{\hat{L}^2}{\hbar^2r^2}
\end{equation}
and
\begin{equation}
\Gamma = \frac{c\hbar}{r}\sqrt{m_0^2c^4 - E^2} + \imath c^2\hbar\frac{(\vec{\sigma} \cdot \vec{r})(\vec{\sigma} \cdot \vec{p})}{r^2}
\end{equation}
It can be seen that eq. (\ref{eq: dirac_so}) reduces to the Klein-Gordon equation of the hydrogen atom on setting $\Gamma=0$. It follows that $\Gamma$ represents the correction needed to take account of the spin on the electron.

Eq. (\ref{eq: conj_dirac_fo_1}) can be written as
\begin{equation} \label{eq: dirac_so_sub}
\tilde{\Phi} = \left(\frac{\lambda}{m_0c^2\hbar\alpha-\lambda\hat{K}}\right)\left( \hbar r\frac{\partial}{\partial r} +\frac{\lambda r}{c} - \frac{\hbar E \alpha}{\lambda } + \hbar \right) \Phi
\end{equation}
giving $\tilde{\Phi}$ in terms of $\Phi$. The remainder of this paper will focus on demonstrating that eqs. (\ref{eq: dirac_so}) and (\ref{eq: dirac_so_sub}) are equivalent to the Dirac equation (\ref{eq: central_dirac}) for the hydrogen atom. In particular, it will be shown that eqs. (\ref{eq: dirac_so}) gives the same energy spectrum as the first order Dirac equation and that the bi-spinor wave functions for hydrogen fall out of eq. (\ref{eq: Psi_a}) and (\ref{eq: Psi_b}).

\section{Solution of the Second Order Equation}
The second order Dirac equation (\ref{eq: dirac_so}) may be expressed as
\begin{equation}
c^2\hbar^2  \left( \frac{\partial^2 }{\partial r^2}+\frac{3}{r}\frac{\partial }{\partial r}-\frac{\hat{K}^2}{\hbar^2r^2} + \frac{1+\alpha^2}{r^2}\right)\Phi + c\hbar \frac{2E\alpha+\lambda}{r}\Phi - \lambda^2 \Phi = 0
\end{equation}
Inserting the product solution 
\begin{equation}
\Phi = \frac{R(r)}{r} \exp\left(-\frac{\lambda r}{c\hbar} \right)\chi_{\kappa m}(\theta, \phi)
\end{equation}
into this gives
\begin{equation} \label{eq: dirac_so_simp_b}
\rho\frac{\partial^2R}{\partial \rho^2}+(1-\rho)\frac{\partial R}{\partial \rho} -\left(\frac{\gamma^2}{\rho}-\frac{\alpha E}{\lambda c \hbar} \right) = 0
\end{equation}
where $\rho=2\lambda r$ and $\gamma = \sqrt{\kappa^2-\alpha^2}$. The substitution
$R = Fr^\gamma$ further reduces eq. (\ref{eq: dirac_so_simp_b}) to the form
\begin{equation}\label{eq: dirac_so_simp_c}
\rho\frac{\partial^2F}{\partial \rho^2}+(b-\rho)\frac{\partial F}{\partial \rho} -aF = 0
\end{equation}
where $a=\gamma-\alpha E / \lambda c \hbar$ and $b=2\gamma+1$. Eq. (\ref{eq: dirac_so_simp_c}) is Kummer's equation \cite{AS}. The physically acceptable solutions of this equation is the confluent hypergeometric function:
\begin{equation}
F(a,b,\rho) = 1 + \frac{a\rho}{b} + \frac{(a)_2\rho^2}{(b)_2 2!} + \ldots +\frac{(a)_n\rho^n}{(b)_n n!} + \dots
\end{equation}
where
\begin{equation}
(a)_n = a(a+1)(a+2) \ldots (a+n-1), (a)_0=1
\end{equation}
Thus, the complete solution of eq. (\ref{eq: dirac_so}) for the hydrogen atom is
\begin{equation} \label{eq: Phi}
\Phi = \exp\left(-\frac{\lambda r}{c\hbar} \right)r^{\gamma-1} F(a, b, \rho)\chi_{\kappa m}(\theta, \phi)
\end{equation}
This is not normalized.

The confluent hypergeometric function $F(a,b,\rho)$ grows exponentially for large $\rho$ unless $a=-n_r$ where $n_r=0, -1, -2, \ldots$. This implies
\begin{equation}
\frac{\alpha E}{c \hbar \lambda} = \gamma + n_r
\end{equation}
or equivalently
\begin{equation} \label{eq: energy_spectrum_a}
E = m_0c^2 \left[ 1 + \frac{\alpha^2}{(n_r+\gamma)^2}\right]^{-1/2}
\end{equation}
For the purpose of correspondence to non-relativistic quantum mechanics, it is usual to define the principal quantum number $n = n_r+|\kappa|$. Eq. (\ref{eq: energy_spectrum_a}) can then be rewritten as
\begin{equation} \label{eq: energy_spectrum_b}
E = m_0c^2 \left[ 1 + \frac{\alpha^2}{(n-|\kappa|+\gamma)^2}\right]^{-1/2}
\end{equation}
It is clear therefore that the first order form of the Dirac equation (\ref{eq: central_dirac}) and the second order form (\ref{eq: dirac_so}) generate the same energy spectrum for the hydrogen atom. 

Inserting eq. (\ref{eq: Phi}) for $\Phi$ into eq. (\ref{eq: dirac_so_sub}) gives
\begin{equation} \label{eq: Phi_conj}
\tilde{\Phi} = \frac{\gamma \lambda - \alpha E}{\alpha m_0c^2 - \kappa \lambda} \exp\left(-\frac{\lambda r}{c\hbar} \right)r^{\gamma-1} F(a+1, b, \rho)\chi_{\kappa m}(\theta, \phi)
\end{equation}
Thus, combining equations (\ref{eq: Psi_a}), (\ref{eq: Psi_b}), (\ref{eq: Phi}) and (\ref{eq: Phi_conj}), the complete bi-spinor eigensolutions of the Dirac equation (\ref{eq: central_dirac}):
\begin{equation}
\Psi_a = \sqrt{m_0c^2+E}{\cal N}_{n\kappa}\exp\left(-\frac{\lambda r}{c\hbar} \right)r^{\gamma-1} [\eta_1 F(a, b, \rho)+\eta_2 F(a+1, b, \rho)]\chi_{\kappa m}(\theta, \phi) 
\end{equation}
\begin{equation}
\Psi_b = \imath \sqrt{m_0c^2-E}{\cal N}_{n\kappa}\exp\left(-\frac{\lambda r}{c\hbar} \right)r^{\gamma-1} [\eta_1 F(a, b, \rho)-\eta_2 F(a+1, b, \rho)]\chi_{-\kappa m}(\theta, \phi) 
\end{equation}
have been constructed for the hydrogen atom. In this, $\eta_1 = \alpha m_0c^2 -\kappa \lambda$, $\eta_2 = \gamma \lambda - \alpha E_{n \kappa}$ and ${\cal N}_{n\kappa}$ is the normalization constant.

\section{Concluding Remarks}
It has been demonstrated the central field Dirac equation can be simplified for, at least, the case of the hydrogen atom using conjugate spinors to substitute for the upper and lower components of the bi-spinor. This leads to a single Klein-Gordon equation for the problem with additional terms to account for the spin of the electron. This approach enables the conjugate spinor wave functions for the hydrogen atom to be calculated alongside the energy spectrum. It has also been shown the conjugate spinor wave functions are readily transformed back into traditional Dirac bi-spinor wave functions.

\end{document}